\documentstyle[preprint,prd,aps]{revtex}

\begin{document}
\draft
\preprint{\ \vbox{
\halign{&##\hfil\cr
       AS-ITP-2000-10\cr\cr}} } \vfil

\title{Exclusive decay of $J/\Psi$ into a lepton pair combined with light hadrons}
\author{J.P. Ma}
\address{Institute of Theoretical Physics,\\
Academia Sinica, \\
P.O.Box 2735, Beijing 100080, China\\
e-mail: majp@itp.ac.cn}
\maketitle

\begin{abstract}
We study the exclusive decay of $J/\Psi$ into a lepton pair combined with
light hadrons in the kinematic region, specified by that the total energy of
the light hadrons is much smaller than $m_c$, the mass of the $c $-quark. In
this region, the nonperturbaive effect related to $J/\Psi$ and that related
to the light hadrons can be separated, the former is represented by a NRQCD
matrix element, while the later is represented by a matrix element of a
correlator of electric chromofields. The results are obtained in a axial
gauge by assumming that contributions from two-gluon emmssion are dominant.
But we can show that these results can be obtained without the
assumption in arbitrary gauges. A discussion of the results are presented.

\end{abstract}

\preprint{\ \vbox{
\halign{&##\hfil\cr
       AS-ITP-2000-10\cr\cr}} } \vfil

\eject
\baselineskip=15pt

\vspace{-5mm}

\vskip20pt \narrowtext
\noindent
{\bf 1. Introduction}

\vskip20pt Several ten millions $J/\Psi$ events will be collected with
Bejing Spectrometer(BES), this provides many opportunities to study
different decay modes of $J/\Psi$ with high statistics. In this work we
propose to study the exclusive decay into a lepton pair combined light
hadrons with a total energy much smaller than the mass of $J/\Psi$, this
decay will be studied experimentally at BES\cite{BES}

Many interesting properties of QCD can be learned from studies of $J/\Psi $
or a quarkonium system. $J/\Psi $ mainly consists of a $c$-quark and its
antiquark and those quarks move with a small velocity $v_c$ inside $J/\Psi $
in its rest-frame. Hence an expansion in $v_c$ can be employed to describe
properties of $J/\Psi $ . The expansion can be systematically performed in
the framework of nonrelativistic QCD(NRQCD)\cite{BBL} for inclusive decays
and for production rates. In this framework the inclusive decay of $J/\Psi $
can be imagined at the leading order of $v_c$ as the following: The $c$- and 
$\bar{c}$- quark in $J/\Psi $ has certain probability to be freed at the
same space point and this $c\bar{c}$ pair decays subsequently. The
probability has a nonperturbative nature and is at order of $v_c^0$, while
the decay of the $c\bar{c}$ pair can be treated with perturbative QCD.
Various effects at higher order of $v_c$ can be taken into account, e.g.,
relativistic effect, and the effect of that the freed $c\bar{c}$ pair does
not possess the same quantum numbers as those of $J/\Psi $.

For the exclusive decay we will study, this interpretation still can be
given at the leading order of $v_c$. Then the decay is mainly of the decay
of the $c\bar{c}$ pair and it can be thought that the pair emit some soft
gluons and annihilates into a virtual photon. The soft gluons will be
transmitted into the light hadrons with a total energy which is small. This
interpretation implies that the nonperturbative effect related to $J/\Psi $
and that related to the light hadrons can be separated, a factorization of
the $S$-matrix element for the decay can be performed. It should be noted
that the factorization here is performed at the amplitude level for the
exclusive decay. It is not like the factorization for inclusive decays
proposed in \cite{BBL}. We will assume that the factorization can be
performed and then try to obtain a factorized form for the decay amplitude.
A proof for the factorization in our case may be done as done for other
exclusive process\cite{Collins}. In our case one may also take the $J/\Psi $
as a nonrelativistic bound-state of a $c\bar{c}$ pair and use a
wave-function for the bound-state. Then the wave-function at the origin can
be written as NRQCD matrix elements defined in \cite{BBL} and our results
can be obtained.

For the light hadrons we will first assume that the soft gluons consists
only of two gluons and take the gauge $G^0(x)=0$, where $G^\mu (x)$ is the
gluon field. Because the soft gluons have smaller momenta compared with the
c-quark mass $m_c$, an expansion in $m_c^{-1}$ can be used. At the leading
order, the S-matrix element for the decay is suppressed by $m_c^{-1}$, this
may be understood in the framework of HQET\cite{HQET}. In the axial gauge
the leading term of HQET does not contain gluons, hence an emission of one
or two gluons is suppressed by $m_c^{-1}$. At the leading order of $m_c^{-1}$
the decay amplitude can be factorized into three parts: The first part is a
NRQCD matrix element representing the bound-state effect of $J/\Psi $, the
second part is a matrix element of a correlator of electric chromofields,
which characterize the soft-gluon transition into the light hadrons. The
third part consists of some coefficient. However, the assumption of
two-gluon-emission is not justified, because it sounds that we use
perturbative theory for soft gluons. But we can show that without the
assumption one can also obtain the same results only by using the expansion
in $m_c^{-1}$. Hence the results are nonperturbative and the number of
emitted gluons is not restricted in an arbitrary gauge. In our work we only
consider the emission of soft gluons which are transmitted into the light
hadrons. Effects of gluons exchanged between quarks, between gluons and
between quarks and gluons are not considered. If the exchanged gluons are
hard, then their effect can be calculated with perturbative theory and it
results in a correction to our results at higher orders of $\alpha _s$. If
the exchanged gluons are soft, their effect is nonperturbative and may be
factorized into nonperturbative matrix elements provided that the 
factorization can be proved. It is beyond this work to
consider the effect of exchanging soft gluons.

Typically, the light hadrons are two pions. Hence the decay studied here has
similarities comparing with the decay $\Psi ^{\prime }\rightarrow J/\Psi
+\pi +\pi $, which is studied in \cite{Pes,Yan,VZ}, where the soft-gluon
transition can be described by matrix elements of local fields in the decay
width. In our case the transition is characterized by a matrix element of
nonlocal fields. However, they are related and we will show this in detail.

Our work is organized as the following: In Sect. 2 we derive our results for
the decay with the assumption of the two-gluon emission in the axial gauge
and discuss the relation to the decay $\Psi ^{\prime }\rightarrow J/\Psi
+\pi +\pi $. In Sect. 3 we will derive the results in arbitrary gauge and
without the assumption, i.e., the number of emitted gluons 
is not restricted. In Sect. 4 we summarize our work and discuss our
approach.

Throughout of our work we take nonrelativistic normalization for the $J/\Psi$
state and for $c$-quark.

\vskip 20pt \noindent
{\bf 2. The Decay with Two-Gluon Emission}

\vskip20pt We consider the exclusive decay of $J/\Psi $ in its rest-frame: 
\begin{equation}
J/\Psi (P)\rightarrow \ell ^{+}(p_1)+\ell ^{-}(p_2)+{\rm light\ hadrons},
\end{equation}
where the momenta are given in the brackets and the light hadrons are
specified hadrons with the total momentum $k$, whose component is much
smaller than $m_c$. Typically, they are two soft pions. At the leading order
of QED the $S$-matrix element for the decay is 
\begin{equation}
\langle f|S|i\rangle =-ie^2Q_cL_\mu \cdot \frac 1{q^2}\int d^4ze^{iq\cdot
z}\langle LH|\bar{c}(z)\gamma ^\mu c(z)|J/\Psi \rangle ,
\end{equation}
where $Q_c$ is the electric charge of $c$-quark in unit of $e$, $c(x)$ is
the Dirac field for $c$-quark, $\langle LH|$ stands for the final state of
the light hadrons and 
\begin{eqnarray}
q &=&p_1+p_2,  \nonumber \\
L_\mu &=&\bar{u}(p_2)\gamma _\mu v(p_1).
\end{eqnarray}
In Eq.(2) we only take the part of $c$-quark in the electric current into
account. Assuming that only two gluons are emitted by the $c$- or $\bar{c}$%
-quark, we obtain 
\begin{eqnarray}
\langle f|S|i\rangle &=&-i\frac 12e^2Q_cg_s^2L_\mu \cdot \frac 1{q^2}\int
d^4xd^4yd^4ze^{iq\cdot z}  \nonumber \\
&&\langle LH|T\left[ \bar{c}(x)\gamma \cdot G(x)c(x)\bar{c}(y)\gamma \cdot
G(y)c(y)\bar{c}(z)\gamma ^\mu c(z)\right] |J/\Psi \rangle ,
\end{eqnarray}
where $G(x)$ is the gluon field. Using Wick-theorem we can calculate the $T$%
-ordered product and we only keep those terms in which one c-field and one $%
\bar{c}$-field remain uncontracted. Then the matrix element takes a
complicated form and can be written in a short notation: 
\begin{eqnarray}
\langle f|S|i\rangle &=&-i\frac 12e^2Q_cg_s^2L_\rho \cdot \frac 1{q^2}\int
d^4xd^4yd^4zd^4x_1d^4y_1 e^{iq\cdot z}\langle LH|G_\mu ^a(x)G_\nu
^b(y)|0\rangle  \nonumber \\
&&\langle 0|\bar{c}_j(x_1)c_i(y_1)|J/\Psi \rangle \cdot M_{ji}^{\mu \nu \rho
,ab}(x,y,x_1,y_1,z),
\end{eqnarray}
where $M_{ji}^{\mu \nu \rho ,ab}(x,y,x_1,y_1,z)$ is a known function, $i$
and $j$ stand for Dirac- and color indices, $a$ and $b$ is the color of
gluon field. The above equation can be generalized to emission of arbitrary
number of soft gluons, then the $S$-matrix element is the sum of the
contributions with 2,3, $\cdots $ soft gluons and in each contribution there
is the same matrix element $\langle 0|\bar{c}_j(x)c_i(y)|J/\Psi \rangle $.
For this matrix element the expansion in $v_c$ can be now performed, the
result is: 
\begin{equation}
\langle 0|\bar{c}_j(x)c_i(y)|J/\Psi \rangle =-\frac 16(P_{+}\gamma ^\ell
P_{-})_{ij}\langle 0|\chi ^{\dagger }\sigma ^\ell \psi |J/\Psi \rangle
e^{-ip\cdot (x+y)}+{\cal O}(v_c^2),
\end{equation}
where $\chi ^{\dagger }(\psi )$ is the NRQCD field for $\bar{c}(c)$ quark
and 
\begin{eqnarray}
P_{\pm } &=&(1\pm \gamma ^0)/2,  \nonumber \\
p^\mu &=&(m_c,0,0,0).
\end{eqnarray}
The leading order of the matrix element is ${\cal O}(v_c^0)$, we will
neglect the contribution from higher orders and the momentum of $J/\Psi $ is
then approximated by $2p$. It should be noted that effects at higher order
of $v_c$ can be added with the expansion in Eq.(6). Taking the result in
Eq.(6) we can write the $S$-matrix element as: 
\begin{eqnarray}
\langle f|S|i\rangle &=&-i\frac 1{24}e^2Q_cg_s^2(2\pi )^4\delta
^4(2p-k-q)L_\rho \cdot \frac 1{q^2}  \nonumber \\
&&\langle 0|\chi ^{\dagger }\sigma ^\ell \psi |J/\Psi \rangle 2^4\int
d^4x\int \frac{d^4q_1}{(2\pi )^4}e^{i2q_1\cdot x}\langle LH|G_\mu ^a(x)G_\nu
^a(-x)|0\rangle R^{\mu \nu \rho \ell }(p,k,q_1).
\end{eqnarray}
To obtain the equation we have used the color-symmetry and the translational
covariance. The quantity $R$ takes the form 
\begin{eqnarray}
R^{\mu \nu \rho \ell }(p,k,q1) &=&{\rm Tr}(P_{+}\gamma ^\ell P_{-})\big\{ %
\gamma ^\rho \frac 1{\gamma \cdot (p-k)-m_c+i0^{+}}\gamma ^\mu \frac 1{%
\gamma \cdot (p-\frac 12k-q_1)-m_c+i0^{+}}\gamma ^\nu  \nonumber \\
&&+\gamma ^\mu \frac 1{\gamma \cdot (-p+\frac 12k-q_1)-m_c+i0^{+}}\gamma ^\nu 
\frac 1{\gamma \cdot (k-p)-m_c+i0^{+}}\gamma ^\rho  \nonumber \\
&&+\gamma ^\mu \frac 1{\gamma \cdot (-p+\frac 12k-q_1)-m_c+i0^{+}}\gamma
^\rho \frac 1{\gamma \cdot (p-\frac 12k-q_1)-m_c+i0^{+}}\gamma ^\nu 
\nonumber \\
&&+(\mu \rightarrow \nu ,\nu \rightarrow \mu ,q_1\rightarrow -q_1)\big\}
\end{eqnarray}
The physical interpretation of $R$ is that it is the amplitude for a $^3S_1$ 
$c\bar{c}$ pair emitting two soft gluons and a virtual photon, and c and $%
\bar{c}$ have the same momentum $p$. We take the gauge 
\begin{equation}
G^0(x)=0,
\end{equation}
and in this gauge the electric chromofield is given by: 
\begin{equation}
{\bf E}(x)={\bf E}^a(x)T^a=\partial _0{\bf G}(x).
\end{equation}
The quantity $R$ depends on $p,k$ and $q_1$. In the kinematic region we
consider, $k$ is a small vector. The dominant region for $q_1$ integration
is characterized by $k$ and by $\Lambda _{QCD}$, because the $x$-dependence
of the matrix element $\langle LH|\cdots |0\rangle $ is characterized by $k$
and by $\Lambda _{QCD}$. Therefore one can expand $R(p,k,q_1)$ in $m_c^{-1}$%
. Keeping only the leading order we obtain 
\begin{eqnarray}
R^{\mu \nu \rho \ell }(p,k,q_1)\langle LH|G_\mu ^a(x)G_\nu ^a(-x)|0\rangle
&=&\frac{4g^{\ell \rho }}{m_ck^0}\cdot \frac 1{k^0+2q_1^0-i0^{+}}\cdot \frac %
1{k^0-2q_1^0-i0^{+}} \cdot((k^0)^2-4(q_1^0)^2)  \nonumber \\
&& \langle LH|{\bf G}^a(x)\cdot {\bf G}^a(-x)|0\rangle +{\cal O}(\frac 1{%
m_c^2})
\end{eqnarray}
Substituting the above expression into the $S$-matrix element, the
integration over ${\bf q_1}$ and over ${\bf x}$ can be performed. Then we
use the following equations 
\begin{eqnarray}
\int d^4xe^{2iq\cdot x}(k+2q)_\rho \langle LH|G_\mu ^a(x)G_\nu
^a(-x)|0\rangle &=&-2i\int d^4xe^{2iq\cdot x}\langle LH|G_\mu ^a(x)
\partial_\rho G_\nu ^a(-x)|0\rangle  \nonumber \\
\int d^4xe^{2iq\cdot x}(k-2q)_\rho \langle LH|G_\mu ^a(x)G_\nu
^a(-x)|0\rangle &=&-2i\int d^4xe^{2iq\cdot x}\langle LH|\partial _\rho G_\mu
^a(x)G_\nu ^a(-x)|0\rangle,
\end{eqnarray}
and we obtain 
\begin{eqnarray}
\langle f|S|i\rangle &=&i\frac{2}{3}e^2Q_cg_s^2(2\pi )^4\delta
^4(2p-k-q)L_\rho \cdot \frac{g^{\rho \ell }}{q^2}\langle 0|\chi ^{\dagger
}\sigma ^\ell \psi |J/\Psi \rangle \cdot \frac 1{m_c}\cdot \frac 1{(k^0)^2} 
\nonumber \\
&&\cdot \int \frac{d\tau}{2\pi} \frac{1}{1+\tau-i0^+}\cdot \frac{1}{%
1-\tau-i0^+} \int dte^{i\tau k^0t}\langle LH|{\bf E}^a(t,{\bf 0})\cdot {\bf E%
}^a(-t,{\bf 0})|0\rangle  \nonumber \\
&& +{\cal O}(\frac 1{m_c^2})+{\cal O}(v_c^2),
\end{eqnarray}
where $\tau$ is related to $q_1^0$ by $2q_1^0=\tau k^0$. The term with $%
g^{\rho \ell }$ is expected in the heavy quark limit. In this limit gluons
will not change the spin of $c$- or $\bar{c}$ quark. The matrix element $%
\langle 0|\chi ^{\dagger }\sigma ^\ell \psi |J/\Psi \rangle $ is
proportional to the spin of $J/\Psi $, hence the spin of $J/\Psi $ is
transferred to the virtual photon. The result is obtained in the axial gauge
defined in Eq.(10), to make the $S$-matrix element gauge invariant we must
add a gauge link between the two operators of the electric chromofield. We
define a distribution amplitude for the gluon conversion into the light
hadrons: 
\begin{equation}
h(\tau) =\frac{g_s^2}{2\pi}\int dte^{i\tau k^0t}\langle LH|{\bf E}^a(t,{\bf 0%
})\cdot \big[ P\exp \{-ig_s\int_{-t}^tdx^0G^{0,c}(x^0,{\bf 0})\tau ^c\}\big]%
_{ab}{\bf E}^b(-t,{\bf 0})|0\rangle ,
\end{equation}
where $P$ means path-ordering and $\tau ^c$ is the generator of $SU(3)$ in
adjoint representation 
\begin{equation}
(\tau ^c)_{ab}=-if_{abc}.
\end{equation}
The function $h(\tau)$ is gauge invariant, it represents the transition
of soft gluons into the light hadrons. With the gauge link the transition is
of two space-like gluon plus many time-like gluons. It also depends on each
individual momentum of each hadron, we denote the dependence as $h(\tau,k)$.
Because of the energy conservation $h(\tau, k) =0$ 
if $\vert \tau\vert >1$. With this the $S$-matrix
element can be written: 
\begin{eqnarray}
\langle f|S|i\rangle &=&i\frac{2}{3}e^2Q_cg_s^2(2\pi )^4\delta
^4(2p-k-q)L_\rho \cdot \frac{g^{\rho \ell }}{q^2}\langle 0|\chi ^{\dagger
}\sigma ^\ell \psi |J/\Psi \rangle \cdot \frac 1{m_c}\cdot \frac 1{(k^0)^2} 
\nonumber \\
&& \cdot T_{LH} (k) +{\cal O}(\frac 1{m_c^2})+{\cal O}(v_c^2),  \nonumber \\
T_{LH}(k)&=&\int d\tau\frac{1}{1+\tau-i0^+}\cdot \frac{1}{1-\tau-i0^+}
h(\tau, k).
\end{eqnarray}
Eq.(15) and Eq.(17) are our main results, derived in the axial gauge with
the assumption of two-gluon emission, then we add a gauge link to make the $%
S $-matrix element gauge invariant. This may be unsatisfied, because the
assumption of two-gluon emission sounds that we performed an expansion in $%
g_s$ for soft-gluons and we add the gauge link by hand. It is possible that
the $c\bar{c}$ pair emits soft gluons which are all time-like and form the
light hadrons, and this emission is not suppressed by $m_c^{-1}$. This type
of contributions is excluded with the gauge. Therefore, it is important to
have the results derived in an arbitrary gauge without the assumption. In
the next section we will show that the results can be obtained without the
assumption and the gauge-link is automatically supplied.

Before ending this section we would like to discuss the relation between the
decay studied here and the decay $\Psi ^{\prime }\rightarrow J/\Psi +\pi
+\pi $. If we expand the integrand in $q_1^0$ in Eq.(12) and then perform
the $q_1^0$ integration we obtain: 
\begin{eqnarray}
\langle f|S|i\rangle &=&i\frac{2}{3}e^2Q_cg_s^2(2\pi )^4\delta
^4(2p-k-q)L_\rho \cdot \frac{g^{\rho \ell }}{q^2}\langle 0|\chi ^{\dagger
}\sigma ^\ell \psi |J/\Psi \rangle \cdot \frac 1{m_c}\cdot \frac 1{(k^0)^3} 
\nonumber \\
&&\sum_{n=0}\frac 1{(k^0)^{2n}}\langle LH|O_{2n}|0\rangle +{\cal O}(\frac 1{%
m_c^2})+{\cal O}(v_c^2),
\end{eqnarray}
where $O_{2n}$ are well known twist-2 operators 
\begin{equation}
O_{2n}=i^{2n}G^{a,0,\mu }({\tensor\partial }_0)^{2n}G^{a,0,\nu }g_{\mu \nu }
\end{equation}
in the axial gauge. With this we can also obtain a series for $T_{LH}(k)$: 
\begin{equation}
T_{LH}(k)=\frac{g_s^2}{k^0}\cdot \sum_{n=0}\frac 1{(k^0)^{2n}}\langle
LH|O_{2n}|0\rangle .
\end{equation}
It should be noted that $T_{LH}$ is defined at a renormalization scale $\mu $%
, the evolution of the operators $O_{2n}$ with $\mu $ is well known, hence
one can also obtain the evolution of $T_{LH}$. The same operators appear in
the decay $\Psi ^{\prime }\rightarrow J/\Psi +\pi +\pi $\cite{Pes,Yan,VZ},
the $S$-matrix element for the decay can be written: 
\begin{equation}
\langle J/\Psi +LH|S|\Psi ^{\prime }\rangle =\sum_{n=0}c_{2n}\langle
LH|O_{2n}|0\rangle ,
\end{equation}
where we denote the state of the two pions as $\langle LH|$. In this decay
the coefficient $c_n$ is proportional to $m_c^{-n}$. Hence one can truncate
the series as an approximation. In our case, the corresponding coefficients
are not suppressed by the power of $m_c^{-1}$, we can only sum the series
into a function $T_{LH}$. This also prevents us from giving numerical
predictions by knowing several leading terms in the series.

\vskip20pt

\noindent
{\bf 3. An exact derivation of the results}

\vskip20pt In this section we work in an arbitrary gauge and do not assume
of two-gluon emission. The number of emitted soft-gluons is not restricted.
With the discussion after Eq.(5) in the last section and after performing
the expansion in $v_c$, the problem can be formulated as a transition of the 
$c\bar{c}$ pair into a virtual photon and the light hadrons. The $S$-matrix
element can be written as 
\begin{eqnarray}
\langle f|S|i\rangle &=&-ie^2Q_cL_\mu \cdot \frac 1{q^2}\frac 16\langle
0|\chi ^{\dagger }\sigma ^\ell \psi |J/\Psi \rangle  \nonumber \\
&&\sum_{s_1s_2}\bar{v}(p,s_2)\gamma ^\ell u(p,s_1)\int d^4ze^{iq\cdot
z}\langle LH|\bar{c}(z)\gamma ^\mu c(z)|c(p,s_1),\bar c(p,s_2)\rangle +{\cal %
O}(v_c^2)  \nonumber \\
&=&-ie^2Q_c(2\pi )^4\delta ^4(2p-k-q)L_\mu \cdot \frac 1{q^2}\frac 16\langle
0|\chi ^{\dagger }\sigma ^\ell \psi |J/\Psi \rangle  \nonumber \\
&&\sum_{s_1s_2}\bar{v}(p,s_2)\gamma ^\ell u(p,s_1)\langle LH|\bar{c}%
(0)\gamma ^\mu c(0)|c(p,s_1),\bar c(p,s_2)\rangle +{\cal O}(v_c^2),
\end{eqnarray}
where the $c$- and $\bar{c}$-quark has same color and the summation over the
color is implied, $s_1$ or $s_2$ is the spin of the $c$-quark or of the $%
\bar c$-quark, respectively. Now we need to calculate the matrix element $%
\langle LH|\bar{c}(0)\gamma ^\mu c(0)|c(p,s_1),\bar c(p,s_2)\rangle $. For
this we use LSZ reduction formula and obtain: 
\begin{eqnarray}
\langle LH|\bar{c}(0)\gamma ^\mu c(0)|c(p,s_1),\bar c(p,s_2)\rangle &=&-%
\frac 1Z\int d^4xd^4ye^{-ip\cdot (x+y)}\bar{v}(p,s_2)(i\gamma \cdot \partial
_x-m_c)  \nonumber \\
&&\cdot\langle LH|T\left\{ c(x)\bar{c}(0)\gamma ^\mu c(0)\bar{c}(y)\right\}
|0\rangle (i\gamma \cdot \overleftarrow{\partial }_y+m_c)u(p,s_1),
\end{eqnarray}
where $Z$ is the renormalization constant for the field $c(x)$. We can
evaluate the matrix element with help of the QCD path-integral, and perform
first the integration over c-quark field we obtain: 
\begin{eqnarray}
\langle LH|\bar{c}(0)\gamma ^\mu c(0)|c(p,s_1),\bar c(p,s_2)\rangle &=&\frac %
1Z\int d^4xd^4ye^{-ip\cdot (x+y)}\bar{v}(p,s_2)(i\gamma \cdot \partial
_x-m_c)  \nonumber \\
&&\langle LH|S(x,0)\gamma ^\mu S(0,y)|0\rangle (i\gamma \cdot \overleftarrow{%
\partial }_y+m_c)u(p,s_1).
\end{eqnarray}
In the above equation the average over gluon fields, i.e., the integration
over gluon fields, will be taken later, and $S(x,y)$ is the c-quark
propagator in a background of gluon fields: 
\begin{equation}
\frac 1iS(x,y)=\langle 0|T\left\{ c(x)\bar{c}(y)\right\} |0\rangle .
\end{equation}
We can define wave-function $\psi _c(x)$ and $\bar{\psi}_{\bar{c}}(x)$ for
the $c$- and $\bar{c}$-quark respectively in a background of gluon fields: 
\begin{eqnarray}
\psi _c(x) &=&\int d^4ye^{-ip\cdot y}\left\{ S(x,y)(i\gamma \cdot 
\overleftarrow{\partial }_y+m_c)\right\} u(p,s_1),  \nonumber \\
\bar{\psi}_{\bar{c}}(x) &=&\int d^4ye^{-ip\cdot y}\bar{v}(p,s_2)(i\gamma
\cdot \partial _y-m_c)S(y,x)
\end{eqnarray}
These wave functions satisfy the Dirac equations 
\begin{eqnarray}
(i\gamma \cdot D-m_c)\psi _c(x) &=&0,  \nonumber \\
\bar{\psi}_{\bar{c}}(x)(i\gamma \cdot \overleftarrow{D}+m_c) &=&0,
\end{eqnarray}
and the boundary conditions 
\begin{eqnarray}
\psi _c(x) &\rightarrow &e^{-ip\cdot x}u(p,s_1)\ \ \ {\rm for}\
x^0\rightarrow -\infty ,  \nonumber \\
\bar{\psi}_{\bar{c}}(x) &\rightarrow &e^{-ip\cdot x}\bar{v}(p,s_2)\ \ {\rm %
for}\ x^0\rightarrow -\infty ,
\end{eqnarray}
where $D^\mu =\partial _\mu +ig_sG_\mu (x)$, $G_\mu (x)$ is the background
field of gluons and $\bar{\psi}\overleftarrow{D}_\mu =(D_\mu \psi )^{\dagger
}\gamma ^0$. It should be noted that the propagator in Eq.(25) is a Feymann
propagator, wave-functions defined in Eq.(26) do satisfy the Dirac equation
with a background field of gluons, but they do not have a simple boundary
condition like in Eq.(28). However, it was shown \cite{Na} that the Feymann
propagator in Eq.(26) can be replaced with a retarded propagator, if the
background field varies enough slowly with the space-time. Then one can have
solutions of Eq.(27) and they satisfy the boundary conditions in Eq.(28). In
our case we will make the expansion in $m_c^{-1}$ and in this expansion the
c-quark and $\bar{c}$ quark are decoupled, and it results in that the
Feymann propagator will automatically be a retarted propagator for c and for 
$\bar{c}$. Therefore one can obtain a solution from the Dirac equation and
the solution can satisfy the boundary condition in Eq.(28). It should be
also noted that the boundary condition in Eq.(28) implies $G^\mu
(x)\rightarrow 0$ for $x^0\rightarrow -\infty $.

With the wave-functions we have 
\begin{equation}
\langle LH \vert \bar c(0) \gamma^\mu c(0) \vert c(p,s_1),\bar c%
(p,s_2)\rangle =\frac{1}{Z} \langle LH \vert \bar\psi_{\bar c}(0) \gamma^\mu
\psi_c(0) \vert 0\rangle
\end{equation}
and it is gauge-invariant. If we take c-quark as a heavy quark, we can
expand the wave-functions in $m_c^{-1}$. The expansion is the same as that
in HQET\cite{HQET}, we write the wave-functions as: 
\begin{eqnarray}
\psi_c(x) &=& e^{-ip\cdot x} \left\{ h_c(x) + \frac{i}{2m_c} {\bf \gamma}%
\cdot D_T h_c +{\cal O}(m_c^{-2})\right\},  \nonumber \\
\bar\psi_{\bar c} (x) &=& e^{-ip\cdot x} \left\{ \bar h_{\bar c} (x) -\frac{i%
}{2m_c} \bar h_{\bar c} (x){\bf \gamma} \cdot \overleftarrow D_T +{\cal O}%
(m_c^{-2})\right\},
\end{eqnarray}
where $D_T^\mu =(0, {\bf D})$. For the functions $h_c(x)$ and $\bar h_{\bar c%
} (x)$ we have 
\begin{eqnarray}
(iD_0 -\frac{1}{2m_c} ({\bf \gamma}\cdot D_T)^2)h_c(x) &=& 0 +{\cal O}%
(m_c^{-2}),  \nonumber \\
\bar h_{\bar c} (x)(i\overleftarrow D_0 -\frac{1}{2m_c} ({\bf \gamma}\cdot 
\overleftarrow D_T)^2) &=& 0 +{\cal O}(m_c^{-2}),
\end{eqnarray}
and the boundary condition for these functions can be read from Eq.(28).
Giving a background field of gluons we solve the above equations by
expanding these functions in $m_c^{-1}$. At the leading order c-quark will
only interact with time-like gluons and the solution is simple. We define 
\begin{equation}
V(x) = P \exp \{-ig_s \int^{x^0}_{-\infty} dt G^0(t,{\bf x})\}
\end{equation}
and with the solution at the leading order the wave-functions are 
\begin{eqnarray}
\psi_c(x) &=& e^{-ip\cdot x} V(x) u(p)+ {\cal O}(m_c^{-1}),  \nonumber \\
\bar\psi_{\bar c} (x)&=& e^{-ip\cdot x} \bar v (p) V^\dagger (x) + {\cal O}%
(m_c^{-1}).
\end{eqnarray}
Because $V^\dagger (x) V(x)=1$ we obtain after averaging the gluon fields: 
\begin{equation}
\langle LH \vert \bar c(0) \gamma^\mu c(0) \vert c(p,s_1),\bar c%
(p,s_2)\rangle =0 +{\cal O}(m_c^{-1}).
\end{equation}
That means that the leading order of the $S$-matrix element is at $m_c^{-1}$
and only time-like gluons will not lead to a contribution to the $S$-matrix
element at order of $m_c^0$. This is in consistency with the results in 
the last section, where the axial gauge and perturbative theory is used. 
Here we derive this and the results at order of $m_c^{-1}$ in an arbitrary 
gauge and without using perturbative theory. 

To calculate the matrix element at order of $m_c^{-1}$, we first make a
gauge transformation: 
\begin{eqnarray}
\psi _c^{\prime }(x) &=&V^{\dagger }(x)\psi _c(x),\ \ \ \bar{\psi}_{\bar{c}%
}^{\prime }(x)=\bar{\psi}_{\bar{c}}(x)V(x),  \nonumber \\
h_c^{\prime }(x) &=&V^{\dagger }(x)h_c^{\prime }(x),\ \ \ \bar{h}_{\bar{c}%
}^{\prime }(x)=\bar{h}_{\bar{c}(x)}V(x),  \nonumber \\
G_\mu ^{\prime }(x) &=&V^{\dagger }(x)G_\mu (x) V(x) -\frac i{g_s}%
V^{\dagger}(x)\partial _\mu V(x).
\end{eqnarray}
After the transformation we have 
\begin{equation}
\langle LH|\bar{c}(0)\gamma^\mu c(0)|c(p,s_1),\bar c(p,s_2)\rangle =\frac 1Z%
\langle LH|\bar{\psi}_{\bar{c}}^{\prime }(0)\gamma ^\mu \psi _c^{\prime
}(0)|0\rangle .
\end{equation}
With this transformation one can show $G^{\prime }{}^0(x)=0$. The wave
functions $\bar\psi _{\bar{c}}^{\prime }$ and $\psi _c^{\prime }$ have the
same expansion in $m_c^{-1}$ as those in Eq.(30), with the replacement that
all functions are replaced by functions with a prime. The functions $%
h_c^{\prime }(x)$ and $\bar{h}_{\bar{c}}^{\prime }(x)$ satisfy the same
equations in Eq.(31), where the covariant derivative is $D_\mu =\partial
_\mu +ig_sG_\mu ^{\prime }(x)$. Solving these equations with the boundary
condition we have 
\begin{eqnarray}
h_c^{\prime }(x) &=&\left\{ 1-\frac i{2m_c}\int_{-\infty }^{x^0}dt(\gamma
\cdot D_T(t,{\bf x}))^2\right\} u(p)+{\cal O}(m_c^{-2}),  \nonumber \\
\bar{h}_{\bar{c}}^{\prime }(x) &=&\bar{v}(p)\left\{ 1-\frac i{2m_c}%
\int_{-\infty }^{x^0}dt(\gamma \cdot D_T(t,{\bf x}))^2\right\} +{\cal O}%
(m_c^{-2}).
\end{eqnarray}
With these solutions we have 
\begin{eqnarray}
\psi _c^{\prime }(x) &=&e^{-ip\cdot x}\left\{ 1-\frac i{2m_c}\int_{-\infty
}^{x^0}dt{\bf G}^{\prime }(t,{\bf x})\cdot {\bf G}^{\prime }(t,{\bf x}%
)+\cdots \right\} u(p)+{\cal O}(m_c^{-2}),  \nonumber \\
\bar{\psi}_{\bar{c}}^{\prime }(x) &=&e^{-ip\cdot x}\bar{v}(p)\left\{ 1-\frac %
i{2m_c}\int_{-\infty }^{x^0}dt{\bf G}^{\prime }(t,{\bf x})\cdot {\bf G}%
^{\prime }(t,{\bf x})+\cdots \right\} +{\cal O}(m_c^{-2}).
\end{eqnarray}
In the above equations $\cdots $ in the bracket $\{\}$ will not lead to
contributions at the order we consider. Using these results we have 
\begin{eqnarray}
B^{\ell\mu } &=&\sum_{s_1,s_2}\bar{v}(p,s_2)\gamma ^\ell u(p,s_1)\langle LH|%
\bar{\psi}_{\bar{c}}(0)\gamma ^\mu \psi _c(0)|0\rangle  \nonumber \\
&=&\frac{-ig_s^2g^{\mu \ell }}{Zm_c}\langle LH|{\bf G}^{\prime }{}^a(0){\bf %
G}^{\prime }{}^a(0)|0\rangle \int_{-\infty }^0dte^{ik^0t}+{\cal O}(m_c^{-2}),
\end{eqnarray}
where we have averaged the gluon fields. It should be noted that $k^0$ is
the total energy of the light hadrons, which forms an asymptotic final state,
hence $k^0$ should be understood as $k^0-i0^{+}$. With this in mind the
integration over $t$ can be performed. One can also use $\theta $-function 
\begin{equation}
\theta (t)=\int_{-\infty }^\infty \frac{d\omega }{2\pi }\frac{ie^{-i\omega t}%
}{\omega +i0^{+}}=\left\{ 
\begin{array}{ll}
1, & t\ge 0 \\ 
0, & t<0
\end{array}
\right.
\end{equation}
to perform the integral. For the transformed gauge fields $G^{\prime }{}^\mu 
$ we have 
\begin{equation}
{\bf E}^{\prime }(x)={\bf E}^{\prime }{}^a(x)T^a=\partial _0{\bf G}^{\prime
}(x),
\end{equation}
then the gluon field ${\bf G}^{\prime }$ can be related to ${\bf E}^{\prime
} $ by 
\begin{equation}
{\bf G}^{\prime }(t,{\bf x})=\int_{-\infty }^td\xi {\bf E}^{\prime }(\xi ,%
{\bf x})=\int_{-\infty }^\infty \theta (t-\xi ){\bf E}^{\prime }(\xi ,{\bf x}%
).
\end{equation}
Using this relation and translational covariance we have 
\begin{eqnarray}
\langle LH|{\bf G}^{\prime }{}^a(0){\bf G}^{\prime }{}^a(0)|0\rangle
&=&2\int_{-\infty }^\infty dtdTe^{ik^0T}\theta (-t-T)\theta (t-T)\langle LH|%
{\bf E}^{\prime }{}^a(t){\bf E}^{\prime }{}^a(-t)|0\rangle  \nonumber \\
&=&2\int_{-\infty }^\infty dt\frac{d\omega }{2\pi }e^{2i\omega t}\cdot \frac %
1{\omega -\frac 12k^0+i0^{+}}\cdot \frac 1{\omega +\frac 12k^0-i0^{+}} 
\nonumber \\
&& \cdot \langle LH|{\bf E}^{\prime }{}^a(t){\bf E}^{\prime
}{}^a(-t)|0\rangle .
\end{eqnarray}
Changing the variable $2\omega=\tau$ and we obtain: 
\begin{eqnarray}
\langle f|S|i\rangle &=&i\frac{2}{3}e^2Q_cg_s^2(2\pi )^4\delta
^4(2p-k-q)L_\rho \cdot \frac{g^{\rho \ell }}{q^2}\langle 0|\chi ^{\dagger
}\sigma ^\ell \psi |J/\Psi \rangle \cdot \frac 1{m_c}\cdot \frac 1{(k^0)^2} 
\nonumber \\
&&\cdot \int \frac{d\tau}{2\pi} \frac{1}{1+\tau-i0^+}\cdot \frac{1}{%
1-\tau+i0^+} \int dte^{i\tau k^0t}\langle LH|{\bf E^{\prime}}^a(t,{\bf 0})
\cdot {\bf E^{\prime}}^a(-t,{\bf 0})|0\rangle  \nonumber \\
&& +{\cal O}(\frac 1{m_c^2})+{\cal O}(v_c^2),
\end{eqnarray}
Expressing ${\bf E}^{\prime }$ with ${\bf E}$ we realize that the gauge link
In Eq.(15) is automatically generated, 
\begin{eqnarray}
&& \langle LH|{\bf E}^{\prime }{}^a(t,{\bf 0})\cdot {\bf E}^{\prime }{}^a(-t,%
{\bf 0})|0\rangle =  \nonumber \\
&& \langle LH|{\bf E}^a(t,{\bf 0})\cdot \big[ P\exp
\{-ig_s\int_{-t}^tdx^0G^{0,c}(x^0,{\bf 0})\tau ^c\}\big]_{ab}{\bf E}^b(-t,%
{\bf 0})|0\rangle
\end{eqnarray}
while in Eq.(15) and Eq.(17) we sandwich it by hand.

The last factor which needs to be considered is the renormalization constant 
$Z$ for wave-function. Following \cite{Na} we consider the current 
\begin{equation}
J^\mu (x)=\bar{c}(x)\gamma ^\mu c(x).
\end{equation}
The current $J^\mu (x)$ is a conserved current and it will be not
renormalized. Hence we have: 
\begin{equation}
\langle c(p)|J^\mu (0)|c(p)\rangle =\bar{u}(p)\gamma ^\mu u(p).
\end{equation}
Using the same techniques in this section, one can derive 
\begin{equation}
Z=1+{\cal O}(m_c^{-1}).
\end{equation}
With the results here we find that the $S$-matrix element obtained here is
exactly the same obtained in the last section. We emphasize here that it is
derived here in arbitrary gauge and without the assumption of two-gluon
emission. Hence the results are nonperturbative.

With the given $S$-matrix element, one can calculate the differential decay
width. We obtain for unpolarized $J/\Psi$: 
\begin{equation}
\frac{d\Gamma}{d q^2 } = \alpha Q_c^2 \frac{8\pi}{9m_c^4} \langle J/\Psi
\vert O_1^{J/\Psi} (^3S_1) \vert J/\Psi \rangle \int
d\Gamma_{LH}\delta((2p-k)^2-q^2)\frac {1}{(k^0)^4} \vert T_{LH} (k) \vert ^2,
\end{equation}
where $\Gamma_{LH}$ is the phase-space integral for the light hadrons and we
only keep the leading dependence on momenta of the light hadrons. The
formula is only for $q^2$ close to $4m_c^2$ and the predicted differential
width is measurable in experiment. In the calculation we used 
\begin{equation}
\langle J/\Psi\vert \psi^\dagger \sigma^i \chi \vert 0 \rangle \langle
0|\chi ^{\dagger }\sigma ^j \psi \vert J/\Psi \rangle = \langle J/\Psi \vert
O_1^{J/\Psi} (^3S_1) \vert J/\Psi \rangle
\varepsilon^i(\varepsilon^j)^\dagger.
\end{equation}
In the equation $\varepsilon$ is the polarization vector of $J/\Psi$, the
matrix element $\langle J/\Psi \vert O_1^{J/\Psi} (^3S_1) \vert J/\Psi
\rangle$ is defined in \cite{BBL} and the average over the spin is implied
in the matrix element.

In this section we mainly have considered emission of gluons by a $c\bar{c}$
pair, and these gluons are transmitted into light hadrons. Because the
kinematic region considered here, these gluons must be soft. That is why we
have defined the nonperturbative object $T_{LH}(k)$ which includes $g_s^2$.
However, emission of hard gluons can happen, this can only happen as
exchange of hard gluons between $c$-, $\bar{c}$- quark and gluons. The
effect of the exchange can be calculated with perturbative QCD, it will
result in that the coefficient in Eq.(17) will be modified. For example, $Z$
in Eq.(48) will be smaller than $1$ by neglecting the effect at $m_c^{-1}$.
For exchange of soft gluons, the effect may be factorized into the NRQCD
matrix element and $T_{LH}$.

\vskip 20pt

\noindent
{\bf 4. Summary and Discussion}

\vskip20pt In this work we have studied the exclusive $J/\Psi $-decay into a
lepton pair combined with light hadrons, in which the total energy of the
light hadrons is much smaller than $M_{J/\Psi }$. The light hadrons are
formed from soft gluons emitted by the $c$-flavored quarks. In the limit $%
m_c\to \infty $ the nonperturbative effects related to $J/\Psi $ and to the
light hadrons can be separated and they are represented by a NRQCD matrix
element and by a correlator of electric chromofields. Corrections to the
limit and those due to hard gluons may systematically added. We emphasize
that our results are obtained without perturbative theory and are gauge invariant,  
although we firstly have derived them in the axial gauge with perturbative 
theory. But we have also derived them in an arbitrary gauge without using 
perturbative theory. 

In our approach an expansion in $v_c$, the velocity of the $c$-quark inside $%
J/\Psi $ in its rest-frame, is used. We have only taken the leading order
contributions at $v_c^0$. At this order, $J/\Psi $ can be considered as a
bound state of the $c\bar{c}$ pair in color singlet. The corrections to the
leading-order results may also added. However, problems arise at order of $%
v_c^4$. At this order $J/\Psi $ has a component in which the $c\bar{c}$ pair
is in color octet and the component is a bound state of the $c\bar{c}$ pair
with soft gluons. It is unclear how to add corrections from this component
at order of $v_c^4$. It deserves a further study of these problems to
understand a bound state of many dynamical freedoms of QCD.

Our approach applies for a class of decays. The results given here can be
directly used for the $\Upsilon $ decay and can be generalized to $%
B_c^{*}\to \ell \bar{\nu}+{\rm light\ hadrons}$, where the light hadrons
also have a total energy which is small. The same correlator appears in
these decays to characterize the transition of soft gluons into light
hadrons. In that sense the correlator is universal. Our results do not apply
for $B_c$, because it is a spin-0 meson. However, with the approach
presented here one may obtain corresponding predictions for $B_c$. Our
results do not apply for decays of excited quaukonium $^3S_1$ states like $%
\Psi ^{\prime }\to \ell ^{+}+\ell ^{-}+\pi +\pi $, in this decay after
emission of soft gluons, which forms the light hadrons, the $c\bar{c}$ pair
can be formed into $J/\Psi $, then the $J/\Psi $ decays into a lepton pair.
This type of contributions is not included in our results. But this type of
contributions may be excluded by experimental cuts, through requiring the
two leptons are not exactly back-to-back in the $J/\Psi $ rest-frame.
With this requirement our results may apply. 

In this work we are unable to give any numerical result. In our prediction,
the nonperturbative effect related to $J/\Psi $ is known, e.g., from the
leptonic decay of $J/\Psi $, while the nonperturbative effect related to the
light hadrons is unknown. We currently develope a model for the correlator
and will make an attempt to give numerical predictions.

\vskip20pt \noindent
{\bf Acknowledgment:} The author would like to thank Prof. Y.Q. Chen, Y.B.
Dai, C. Liu, D. Soper and Z.X. Zhang for interesting discussions. This work
is supported by National Science Foundation of P.R. China and by the Hundred
Yonng Scientist Program of Sinica Academia of P.R.China.

\vskip15pt

\vfil\eject


\begin{references}
\bibitem{BES}  Z. G. Zhao, BES Collabration, private communication

\bibitem{BBL}  G.T Bodwin, E. Braaten, and, G.P Lepage, Phys. Rev. D51
(1995) 1125

\bibitem{HQET}  N. Isgur and M.B. Wise, Phys. Lett. B232 (1989) 113, ibid
B237 (1990) 527

E. Eichten and B. Hill, Phys. Lett. B234 (1990) 511

B. Grinstein, Nucl. Phys. B339 (1990) 253

H. Georgi, Phys. Lett. B240 (1990) 447

\bibitem{Collins}  J.C. Collins, L. Frankfurt and M. Strikman, Phys. Rev.
D56 (1997) 2982

J.C. Collins and A. Freund, Phys. Rev. D59 (1999) 074009

\bibitem{Pes}  M.E. Peskin, Nucl. Phys. B156 (1979) 365,

G. Bhanot and M.E. Peskin, Nucl. Phys. B156 (1979) 391

\bibitem{Yan}  T.M. Yan, Phys. Rev. D22 (1980) 1652

\bibitem{VZ}  M. Voloshin and V. Zakharov, Phys. Rev. Lett. Vol. 45 (1980)
688

\bibitem{Na}  O. Nachtmann, Ann. Phys. Vol. 209 (1991) 436
\end{references}
\end{document}